# Nonsingular two dimensional cloak of arbitrary shape

Jin Hu, Xiaoming Zhou and Gengkai Hu[*]

*School of Aerospace Engineering, Beijing Institute of Technology, Beijing 100081,*

*People's Republic of China.*

[*]*Corresponding author: hugeng@bit.edu.cn*

We propose a general method to circumvent the singularity of arbitrary 2D cloaks, which arises from infinitely large values of material parameters at inner boundaries. The presented method is based on the deformation view of the transformation design method. It is shown that by adjusting the principle stretch out of the cloaking plane, 2D cloaks of arbitrary shapes without singularity can be constructed. It is also demonstrated that the method based on the equivalent dispersion relation and the design method for nonsingular 2D cloak from mirror-symmetric cross section of 3D cloak can be derived from the proposed theory. Examples of a cylindrical electromagnetic cloak and an arbitrary shaped 2D electromagnetic cloak without singularity are provided to illustrate the method.

Recently, there have been intensive studies on the transformation optics [1-3], a general method that enables one to design a device with prescribed functionality, for example the invisibility cloaks [2,4]. The designed cloaks are locally anisotropic and spatially inhomogeneous. For a two dimensional (2D) cloak, the necessary material parameters at the inner boundary is usually singular [5], this singular problem significantly limits the practical realization of cloaks. For carpet or cylindrical cloaks, the singularity can be avoided by introducing special transformations, such as by expanding the cloak region from a line segment [6,7], instead of a point. Another interesting method [8] designs a nonsingular 2D cloak by projecting on a mirror-symmetric cross section from a 3D cloak, which is inherently nonsingular [5]. One can also design approximately a 2D cloak without singularity by carefully tuning the material parameters under the condition of the equivalent dispersion relation as an ideal singular one [4,9-11]. Design with nonsingular or simpler material parameters for a given functionality is an important step for real engineering applications, especially for broadband applications [4,6-12]. However to our knowledge, there is not a general method that can design nonsingular 2D cloaks of random shapes, so the objective of this letter is to propose one. The proposed method is based on the deformation view of the transformation method recently established by Hu *et al*. [13]. Numerical examples will be given to demonstrate how a nonsingular arbitrary cloak can be constructed.

Let us firstly explain how the singularity is formed during the construction of a 2D cloak. Transformation optics is based on form-invariant Maxwell's equation during coordinate transformation. The spatial coordinate transformation from a flat space **x** to a distorted space **x**'(**x**) is equivalent to material parameter variations in the original flat space. The permittivity $\varepsilon'$ and permeability $\mu'$ in the transformed space are given by [5]

$$\boldsymbol{\varepsilon}' = \mathbf{A}\boldsymbol{\varepsilon}_0\mathbf{A}^{\mathrm{T}}/\det\mathbf{A}, \tag{1a}$$

$$\boldsymbol{\mu}' = \mathbf{A}\boldsymbol{\mu}_0\mathbf{A}^{\mathrm{T}}/\det\mathbf{A}, \tag{1b}$$

where $\mathbf{A}$ is the Jacobian transformation tensor with components $A_{ij}=\partial x'_i/\partial x_j$. From the deformation perspective [13] of coordinate grids, the material parameters $\boldsymbol{\varepsilon}'$ and $\boldsymbol{\mu}'$ are related directly to the pure stretch component of the deformation gradient $\nabla\mathbf{x}'$ that characterizes the distortion of the original flat grids. Suppose the principal stretches [14] are denoted by $\lambda_1$, $\lambda_2$, and $\lambda_3$ respectively in the principal directions and for the original material $\boldsymbol{\varepsilon}_0 = \boldsymbol{\mu}_0 = 1$. In the same principal system, equation (1) can be rewritten as the following diagonal form [13]

$$\boldsymbol{\varepsilon}' = \boldsymbol{\mu}' = \begin{bmatrix} \dfrac{\lambda_1}{\lambda_2\lambda_3} & 0 & 0 \\ 0 & \dfrac{\lambda_2}{\lambda_1\lambda_3} & 0 \\ 0 & 0 & \dfrac{\lambda_3}{\lambda_1\lambda_2} \end{bmatrix}. \tag{2}$$

In order to make a perfect cloak, its outer boundary needs to be fixed, i.e. $\mathbf{x}'=\mathbf{x}$ [5]. This condition naturally constraints the stretches to unity (without deformation) in the tangential directions at the outer boundary. The perfectly matched layer (PML) of the outer boundary has additional conditions: the unit tangential stretches are the corresponding principal stretches, the third principal stretch must be normal to the boundary and has a value equal to the component of the transformed material parameters in that direction, i.e., $\lambda_n = a_n, \lambda_{t1}=1, \lambda_{t2}=1$ [15].

For a cylindrical cloak [4], the material parameters and principal stretches of the transformation-induced deformation are sketched in Fig. 1. For a linear transformation $r' = a + \dfrac{b-a}{b}r$, $\theta' = \theta$ and the cloak is bordered by $r' \in (a,b)$, the principal stretches of each point within the cloak are given by

$$\lambda_r = \frac{dr'}{dr} = \frac{b-a}{b}, \tag{3a}$$

$$\lambda_\theta = \frac{r'd\theta'}{rd\theta} = \frac{r'}{r'-a}\frac{b-a}{b}, \tag{3b}$$

$$\lambda_z = \frac{dz'}{dz} = 1, \tag{3c}$$

where no transformation is assumed in the $z$ direction, i.e. $z' = z$. At the inner boundary $r' = a$ of the cloak, it can be seen from Eq. (3) that $\lambda_\theta \to \infty$, while $\lambda_r$ and $\lambda_z$ are kept finite. Due to the infinite stretch $\lambda_\theta$ in the azimuthal direction, some components in permittivity and permeability tensors may approach infinite values near the inner boundary, as seen in Eq.(2).

For a 2D cloak, the stretch $\lambda_z$ perpendicular to the cloaking plane is decoupled with the in-plane stretches $\lambda_r$ and $\lambda_\theta$, so the continuously variant stretch $\lambda_z$ can be chosen freely except that $\lambda_z = 1$ is deeded at the outer boundary in order to satisfy the impedance-matching condition. Hence the singularity can be avoided if we set $\lambda_z = \lambda_\theta$. In this case, the ratios of the stretches in Eq. (2) are finite. From Eq. 3(b), it can be found that $\lambda_z = 1$ is satisfied at the outer boundary. According to Eq. (2), we get a nonsingular perfect cylindrical cloak as

$$\varepsilon'_r = \mu'_r = \frac{\lambda_r}{\lambda_\theta \lambda_z} = \frac{b}{b-a}\left(\frac{r'-a}{r'}\right)^2, \tag{4a}$$

$$\varepsilon'_\theta = \mu'_\theta = \frac{\lambda_\theta}{\lambda_z \lambda_r} = \frac{b}{b-a}, \tag{4b}$$

$$\varepsilon'_z = \mu'_z = \frac{\lambda_z}{\lambda_r \lambda_\theta} = \frac{b}{b-a}. \tag{4c}$$

The material parameters in Eq. (4) are exactly the same as those reported in references [10,11] based on a different method.

To generalize the above idea to 2D cloaks of arbitrary shape, we observe that the in-plane ($x_1, x_2$) and out-of-plane ($x_3$) deformations are decoupled. According to Hu *et al.* [13], the in-plane deformation can be solved from the Laplace's equation with proper boundary conditions

$$\left(\frac{\partial^2}{\partial x_1'^2} + \frac{\partial^2}{\partial x_2'^2}\right) x_i = 0, \; i=1,2, \tag{5a}$$

$$\mathbf{x}\big|_{\mathbf{x}' \in \partial\Omega_+} = \mathbf{x}', \; \mathbf{x}\big|_{\mathbf{x}' \in \partial\Omega_-} = 0, \tag{5b}$$

where $\partial\Omega_+$ and $\partial\Omega_-$ are the outer and inner boundaries of the 2D cloak respectively. For the out-of-plane stretch, $x_3' = x_3$ is assumed. Note that $\lambda_3 = 1$ is satisfied at the outer boundary. The deformation gradient tensor $\mathbf{A} = \nabla \mathbf{x}'$ can be inversely obtained from Eq. (5), as well as the left Cauchy-Green deformation tensor $\mathbf{B} = \mathbf{A}\mathbf{A}^\mathrm{T}$. In Cartesian coordinate system, we have

$$\mathbf{A} = \begin{bmatrix} A_{11} & A_{12} & 0 \\ A_{21} & A_{22} & 0 \\ 0 & 0 & 1 \end{bmatrix}, \; \mathbf{B} = \begin{bmatrix} B_{11} & B_{12} & 0 \\ B_{12} & B_{22} & 0 \\ 0 & 0 & 1 \end{bmatrix}. \tag{6}$$

The in-plane principal stretches can be calculated directly from $\mathbf{B}$ as

$$\lambda_{1,2} = \sqrt{\frac{B_{11} + B_{22} \mp \sqrt{B_{11}^2 - 2B_{11}B_{22} + B_{22}^2 + 4B_{12}^2}}{2}}. \tag{7}$$

$\lambda_2$ is always greater than $\lambda_1$ and it will tend to infinite near the inner boundary. To avoid the singularity, we let $\lambda_3$ tend to infinity with same order as $\lambda_2$ at the inner boundary and $\lambda_3 = 1$ at the outer boundary. For an arbitrary 2D cloak, we can not simply set $\lambda_3 = \lambda_2$, since there may be $\lambda_2 \neq 1$ at the outer boundary, which means the principle stretch may not be tangential to the outer boundary. For this reason, we choose the out-of-plane stretch $\lambda_3$ as

$$\tilde{\lambda}_3 = C_0(|x'_1 - x_1| + |x'_2 - x_2|)\lambda_2 + 1. \tag{8}$$

where $C_0$ is a constant value. Then Eq. (6) can be rewritten as

$$\tilde{\mathbf{A}} = \begin{bmatrix} A_{11} & A_{12} & 0 \\ A_{21} & A_{22} & 0 \\ 0 & 0 & \tilde{\lambda}_3 \end{bmatrix}, \tilde{\mathbf{B}} = \begin{bmatrix} B_{11} & B_{12} & 0 \\ B_{12} & B_{22} & 0 \\ 0 & 0 & \tilde{\lambda}_3^2 \end{bmatrix}. \tag{9}$$

For an arbitrary 2D cloak, the permittivity and permeability tensors in the transformed space are finally given by [13]

$$\boldsymbol{\varepsilon}' = \boldsymbol{\mu}' = \tilde{\mathbf{B}} / \det \tilde{\mathbf{A}}. \tag{10}$$

As discussed above, the designed 2D cloak will have no singular material parameter and the outer boundary is impedance matching. Theoretically, there are no reflections at inner boundary, and no waves can penetrate into the cloaked region [5, Sec. 5.2]. As an example of an arbitrary cloak illuminated by plane harmonic waves, the simulation results of an arbitrary shaped 2D cloak designed by the proposed method are given in Fig. 2, where $C_0 = 5$ is taken in Eq. (8). The corresponding material parameters are shown in Fig. 3. As shown in these figures, the cloaking effect can really be achieved by finite material parameters.

The coordinate transformation for the out-of-plane displacement $x'_3$ can be calculated from $\tilde{\lambda}_3$. In Cartesian coordinate system, it reads $x'_3 = \tilde{\lambda}_3(x'_1, x'_2)x_3 + C_1$, where $C_1$ is a constant value. Usually the corresponding spatial transformations for $\tilde{\lambda}_3$ are not unique; in order to construct the nonsingular 2D cloak, we can associate it with a 3D cloak constructed with a special transformation perpendicular to the cloaking plane, so the proposed construction for 2D nonsingular cloak can be considered as the projection on the cloaking plane from a 3D cloak, which is constructed by expending a point into a finite boundary through a special transformation, while the outer boundary is fixed. An example of the nonsingular cylindrical cloak constructing is shown in Fig. 4. In this sense, the method in Ref. [8] for designing 2D cloak with complex shapes from the mirror-symmetric cross section of 3D cloak can be understood.

Now recall the cylindrical cloak, the stretch perpendicular to the cloaking plane can be chosen freely. Then we choose $\lambda_z = r'/(r'-a)$ instead of $\lambda_z = \lambda_\theta = r'(b-a)/[(r'-a)b]$, leading to the following parameters for the cloak

$$\varepsilon'_r = \mu'_r = \frac{\lambda_r}{\lambda_\theta \lambda_z} = (\frac{r'-a}{r'})^2, \tag{11a}$$

$$\varepsilon'_\theta = \mu'_\theta = \frac{\lambda_\theta}{\lambda_z \lambda_r} = (\frac{b}{b-a})^2, \tag{11b}$$

$$\varepsilon'_z = \mu'_z = \frac{\lambda_z}{\lambda_r \lambda_\theta} = 1. \tag{11c}$$

The result in Eq. (11) is the same as that derived by Schurig *et al.* and Cummer *et al.* [4,9] by the equivalent dispersion relations. However at the outer boundary, $\lambda_z = b/(b-a)$ violates the impedance-matching condition $\lambda_z = 1$. So the designed cloak will have nonzero reflectance [4,9].

In conclusion, the geometrical view of Eq. (2) on transformation optics can be used to simplify material parameters to avoid the singularity in designing 2D effective cloaks of arbitrary shape. It is shown that the existing methods such as projection method and the method based on equivalent dispersion relation can be better understood from the present method. Full-wave simulations validate the design method for a 2D cloak of arbitrary shape without singularity. Since the proposed method is completely based on the spatial transformation without other additional assumptions, it can be extended to the acoustic case [16,17] directly.

This work is supported by the National Natural Science Foundation of China (90605001, 10702006, 10832002), and the National Basic Research Program of China (2006CB601204).

Figure captions:

Figure 1. (Color online) Sketch of the material parameters and principal stretches in the transformed space of a cylindrical cloak.

Figure 2. (Color online) The contour plots of the electric field $E_z$ for the TE waves incident from (a) left and (b) bottom on a irregular shaped cloak without singularity. The white lines indicate directions of the power flow.

Figure 3. (Color online) The contour plots of the material parameters (a) $\varepsilon_z$, (b) $\mu_{xx}$, (c) $\mu_{xy}$ and (d) $\mu_{yy}$ of the irregular cloak with $C_0$=5 in Eq. (8).

Figure 4. (Color online) The great-circle cross section of the spherical cloak that only has transformation in the radial direction. Stretch $\lambda_z$ locally perpendicular to the cloak plane can be found to equal to $\lambda_\theta$ that should be infinite at the inner boundary in the cloak plane.

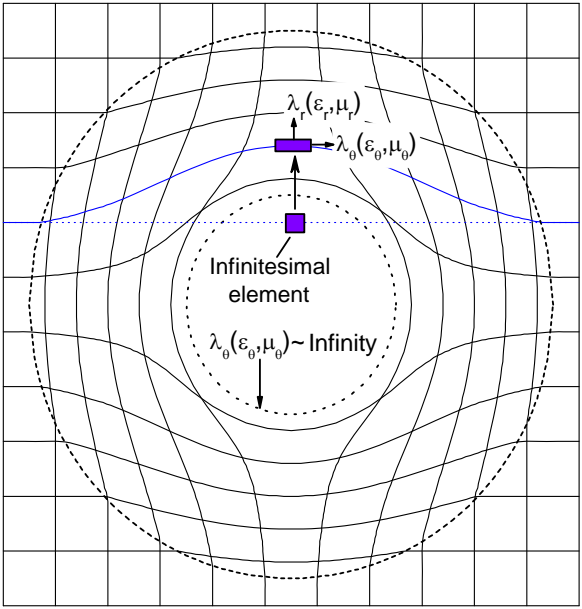

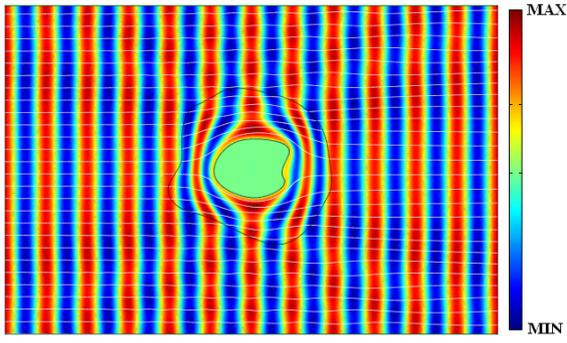 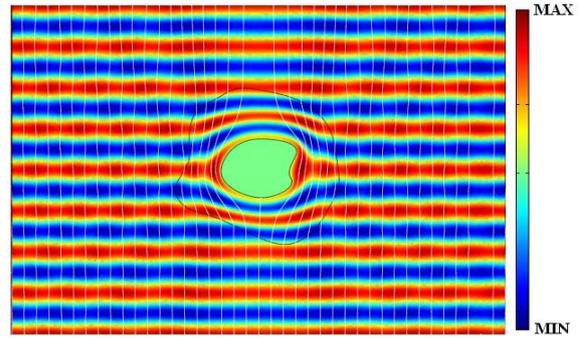

(a) (b)

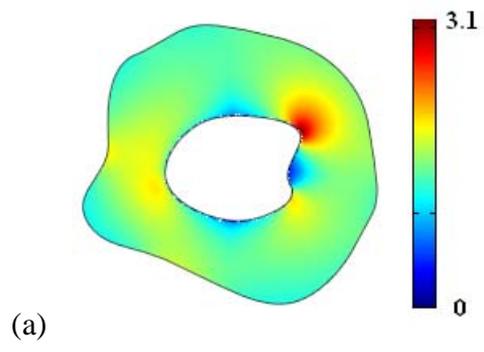 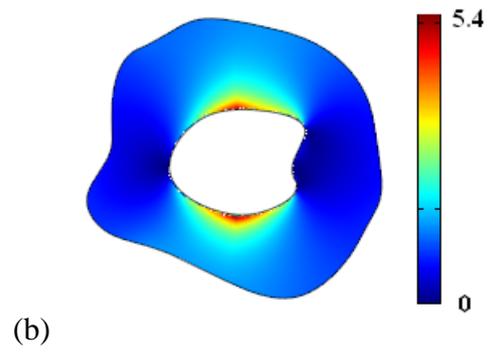

(a) (b)

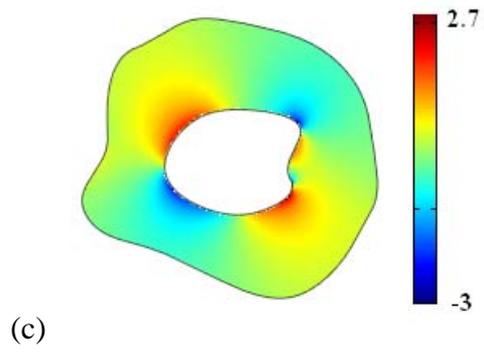 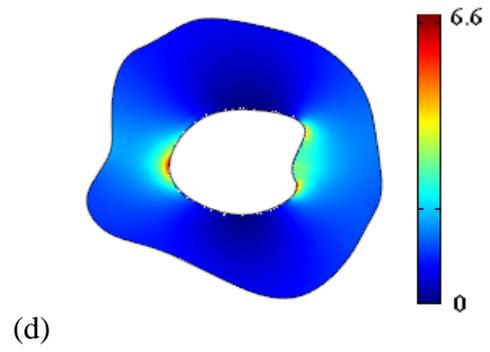

(c) (d)

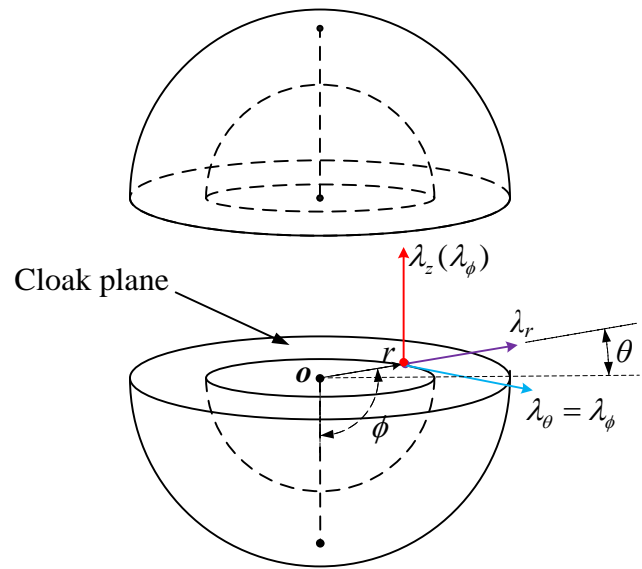